\documentclass[aps,prb,twocolumn,superscriptaddress]{revtex4}
\usepackage{epsfig}

\renewcommand{\thefootnote}{\fnsymbol{footnote}}
\renewcommand{\theequation}{\arabic{section}.\arabic{equation}}

\usepackage{amssymb}
\usepackage{amsfonts}
\usepackage{amsmath}
\usepackage{bm}
\usepackage{textcomp}
\usepackage{color}
\usepackage{longtable}
\usepackage{graphicx}
\usepackage{dcolumn}

\usepackage{epsfig}

\begin{document}
\renewcommand{\thefootnote}{\fnsymbol{footnote}}
\renewcommand{\theequation}{\arabic{section}.\arabic{equation}}

\title{Experimental probing of exchange interactions between localized spins in the dilute magnetic insulator (Ga,Mn)N}

\author{A. Bonanni} \email{alberta.bonanni@jku.at}
\affiliation{Institut f\"ur Halbleiter- und Festk\"orperphysik,
  Johannes Kepler University, Altenbergerstr.~69, 4040 Linz,
  Austria}

\author{M. Sawicki} \email{mikes@ifpan.edu.pl}
\affiliation{Institute of Physics, Polish Academy of Sciences,
  al.~Lotnik\'ow 32/46, 02-668 Warszawa, Poland}

\author{T. Devillers} \affiliation{Institut f\"ur Halbleiter- und
  Festk\"orperphysik, Johannes Kepler University, Altenbergerstr.~69,
 4040 Linz, Austria}

\author{W. Stefanowicz}
\affiliation{Institute of Physics, Polish Academy of Sciences,
  al.~Lotnik\'ow 32/46, 02-668 Warszawa, Poland}
\affiliation{Laboratory of Magnetism,
  Bialystok University, ul. Lipowa 41, 15-424 Bialystok, Poland}

\author{B. Faina} \affiliation{Institut f\"ur Halbleiter- und
  Festk\"orperphysik, Johannes Kepler University, Altenbergerstr.~69,
  4040 Linz, Austria}

\author{Tian Li} \affiliation{Institut f\"ur Halbleiter- und
  Festk\"orperphysik, Johannes Kepler University, Altenbergerstr.~69,
4040 Linz, Austria}

\author{T. E. Winkler} \affiliation{Institut f\"ur Halbleiter- und
  Festk\"orperphysik, Johannes Kepler University, Altenbergerstr.~69,
 4040 Linz, Austria}

\author{D.~Sztenkiel} \affiliation{Institute of Physics, Polish
  Academy of Sciences, al.~Lotnik\'ow 32/46, 02-668 Warszawa,
  Poland}

\author{A. Navarro-Quezada} \affiliation{Institut f\"ur
  Halbleiter- und Festk\"orperphysik, Johannes Kepler University,
  Altenbergerstr.~69, 4040 Linz, Austria}

\author{M. Rovezzi} \affiliation{Institut f\"ur
  Halbleiter- und Festk\"orperphysik, Johannes Kepler University,
  Altenbergerstr.~69, 4040 Linz, Austria}

\author{R. Jakie\l a} \affiliation{Institute of Physics,
  Polish Academy of Sciences, al.~Lotnik\'ow 32/46, 02-668 Warszawa,
  Poland}

\author{A. Grois} \affiliation{Institut f\"ur
  Halbleiter- und Festk\"orperphysik, Johannes Kepler University, Altenbergerstr.~69, 4040 Linz, Austria}

\author{M. Wegscheider} \affiliation{Institut f\"ur
  Halbleiter- und Festk\"orperphysik, Johannes Kepler University, Altenbergerstr.~69, 4040 Linz, Austria}

\author{W. Jantsch} \affiliation{Institut f\"ur
  Halbleiter- und Festk\"orperphysik, Johannes Kepler University, Altenbergerstr.~69, 4040 Linz, Austria}

\author{J. Suf\mbox{}fczy\'{n}ski}
\affiliation{Faculty of Physics, University of Warsaw,
 00-681 Warszawa, Poland}

 \author{F. D'Acapito}
\affiliation{CNR-INFM-OGG, Italian Collaborating Research Group, "GILDA" - ESRF, 38043 Grenoble, France}

\author{A. Meingast} \affiliation{Institute for Electron Microscopy -- FELMI,
  Graz University of Technology, 8010 Graz, Austria}

\author{G. Kothleitner} \affiliation{Institute for Electron Microscopy -- FELMI,
  Graz University of Technology, 8010 Graz, Austria}

\author{T. Dietl} \email{dietl@ifpan.edu.pl}
\affiliation{Institute of Physics, Polish Academy of Sciences,
  al.~Lotnik\'ow 32/46, 02-668 Warszawa, Poland}
\affiliation{Faculty of Physics, University of Warsaw,
 00-681 Warszawa, Poland}

\begin{abstract}
The sign, magnitude, and range of the exchange couplings between pairs of Mn ions is determined for (Ga,Mn)N and (Ga,Mn)N:Si with $x \lesssim 3$\%. The samples have been grown by metalorganic vapor phase epitaxy and characterized by secondary-ion mass spectroscopy; high-resolution transmission electron microscopy with capabilities allowing for chemical analysis, including the annular dark-field mode and electron energy loss spectroscopy;  high-resolution and synchrotron x-ray diffraction; synchrotron extended x-ray absorption fine-structure; synchrotron x-ray absorption near-edge structure; infra-red optics and electron spin resonance. The results of high resolution magnetic measurements and their quantitative interpretation have allowed to verify a series of {\em ab initio} predictions on the possibility of ferromagnetism in dilute magnetic insulators and to demonstrate that the interaction changes from ferromagnetic to antiferromagnetic when the charge state of the Mn ions is reduced from 3+ to 2+.

\end{abstract}

\date{\today}

\maketitle

\section{Introduction}
\begin{figure}[tb]  
  \centering
  \includegraphics[width=8cm]{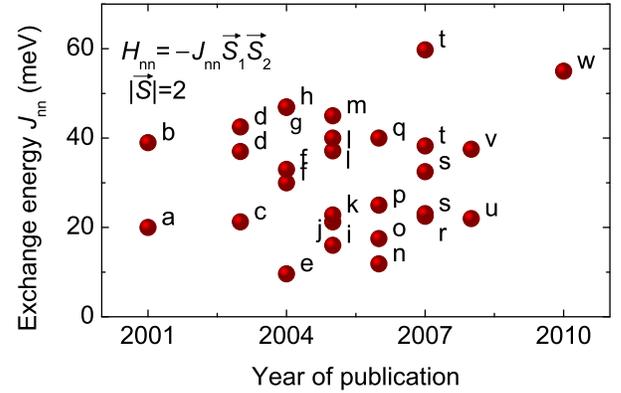}
  \caption{(Color online) Exchange energy $J_{\text{nn}}$ for the nearest neighbor (nn) coupling from {\em ab initio} computations by various authors, as listed in Ref.~\onlinecite{Sato:2001_JJAP}. To obtain $J_{\text{nn}}$, the calculated energy differences for antiferromagnetic and ferromagnetic arrangements of spins within the nn Mn pair of the determined exchange energies have been mapped on the classical Heisenberg Hamiltonian.}
  \label{fig:PRB_JNN}
\end{figure}

The decisive role of holes in ordering the localized spins in dilute magnetic semiconductors (DMSs) is not only well established,\cite{Dietl:2000_S,Jungwirth:2006_RMP,Sato:2010_RMP} but it represents also the basis of the functionalities demonstrated for these systems.\cite{Spintronics:2008_B} In view of the fact that most magnetic insulators are either antiferromagnets or ferrimagnets, particularly intriguing is the question whether ferromagnetism is at all possible in dilute magnetic insulators, where carriers remain strongly localized on parent impurities or defects.\cite{Sato:2010_RMP,Zunger:2010_P,Dietl:2008_PRB} Actually, a {\em ferromagnetic} coupling between localized spins was predicted in a series of {\em ab initio} works for the model system (Ga,Mn)N,\cite{Sato:2001_JJAP} where, as shown in Fig.~\ref{fig:PRB_JNN}, a large value of the ferromagnetic exchange energy $J_{\text{nn}}$ was calculated for the nearest neighbor (nn) Mn pairs. Due to the highly localized character of the orbitals in question, the magnitude of $J$ is expected to decay fast with the inter--spin distance. Nevertheless, according to recent Monte Carlo simulations, the predicted Curie temperature $T_{\mathrm{C}}$ is as high as 35~K and 65~K for the Mn cation concentration $x=3$\% and 6\%, respectively.\cite{Sato:2010_RMP}

As reviewed elsewhere,\cite{Liu:2005_JMSME,Bonanni:2007_SST,Dietl:2010_NM}  this clear-cut theoretical prediction has not been yet verified experimentally. Instead, a diversity of magnetic properties has been reported. For instance,  no indication of ferromagnetic interactions was detected up to $x = 36$\% in polycrystalline films prepared by ion-assisted deposition,\cite{Granville:2010_PRB}   whereas $T_{\mathrm{C}}$ values ranging from 8~K (Ref.~\onlinecite{Sarigiannidou:2006_PRB}) up to over 300~K (Ref.~\onlinecite{Liu:2005_JMSME}) were found for (Ga,Mn)N grown by molecular beam epitaxy (MBE). However, the detection of phase separations\cite{Zajac:2003_JAP,Dhar:2003_APL,Martinez-Criado:2005_APL,Shon:2008_MSEB} may suggest that the determined $T_{\mathrm{C}}$ corresponds to the blocking temperature of magnetic nanoparticles. It is increasingly clear that a further progress in the understanding of this challenging system requires a precise control of both the spatial distribution and the charge state of the Mn ions.\cite{Bonanni:2007_SST}

The samples whose properties are discussed in this paper have been grown by matalorganic vapor phase epitaxy (MOVPE) according to a procedure described in Sec.~II. As detailed in Sec.~III, our films have been characterized by: secondary-ion mass spectroscopy (SIMS); high-resolution (scanning) transmission electron microscopy [HR-(S)TEM] with capabilities allowing for chemical analysis, including the energy dispersive x-ray spectroscopy (EDS), high angle annular dark-field (HAADF) mode, and electron energy loss spectroscopy (EELS);  high-resolution and synchrotron x-ray diffraction (SXRD); synchrotron extended x-ray absorption fine-structure (EXAFS); synchrotron x-ray absorption near-edge structure (XANES); infra-red optics and electron spin resonance (ESR). This set of probes demonstrates the absence of precipitation, reveals a non-uniform  Mn distribution in the digitally Mn-doped films, and shows that the concentration of Mn$^{2+}$ ions reaches $4\times 10^{20}$~cm$^{-3}$ in (Ga,Mn)N:Si. From four probe conductance measurements, the sheet resistance is of the order of 10~G$\Omega$ at 300~K for the (Ga,Mn)N film with the
highest Mn content $x = 3.1$\%, where $x$ is the concentration of Mn cations.

By combining this extensive growth and nanocharacterization program with the results of high-precision magnetic measurements discussed in Sec.~IV we demonstrate that the dominant interactions between neighbor Mn pairs are ferromagnetic in (Ga,Mn)N. However, according to the data, the coupling is too short-ranged to lead to magnetic ordering above 1.85~K in the studied Mn concentration range up to $x = 3$\%. Employing a model of magnetic susceptibility suitable for wurtzite (Ga,Mn)N at high temperatures, and outlined in the Appendix, we evaluate from our experimental results the magnitude of the exchange energy for the nearest neighbor ferromagnetic coupling.   These findings allow to verify the series of {\em ab initio} predictions summarized in Fig.~\ref{fig:PRB_JNN} on the possibility of ferromagnetism in dilute magnetic insulators. At the same time, we show that the interactions become antiferromagnetic if the Mn charge state is altered by co-doping with Si donors, clarifying in this way the array of magnetic properties reported for this system.

\section{Growth method and studied samples}

In order to increase the Mn concentration in (Ga,Mn)N grown on GaN/$c$-sapphire by MOVPE at a substrate temperature of 850$^{\circ}$C,\cite{Stefanowicz:2010_PRBa} here the flow--rate of the Ga precursor (TMGa) is reduced to 1 standard cubic centimeter per minute (sccm), maintaining the temperature of the Mn precursor source (MeCp$_{2}$Mn) at 22$^{\circ}$C and its flow rate up to 490~sccm. In addition to the uniformly doped Mn films, we also grow digitally ($\delta$) Mn-doped structures, in which the Mn and Ga precursors are supplied alternately with a period ratio up to 8. Furthermore, a series of respectively uniformly and digitally Mn-doped samples is co-doped with Si at a SiH$_4$ flow rate of either 1 or 2~sccm. The four types of considered samples are denoted by (Ga,Mn)N, (Ga,$\delta$Mn)N, (Ga,Mn)N:Si, and (Ga,$\delta$Mn)N:Si respectively, where with $\delta$Mn we refer to the Mn-digitally doped layers.

In Table~I the list of the studied samples is presented, whose magnetic properties are  reported in Figs.~\ref{fig:PRB_M_H} and \ref{fig:PRB_M_H_Si} (Sec.~IV). An additional series of samples has been grown onto double-side epi-ready substrates for optical transmission studies. The Mn concentration $x$ ($x_{\mathrm{av}}$ for the digital structures), as determined by the near-saturation value of the in-plane magnetization $M$ at 50~kOe and 1.85~K, reaches over 3\% in the samples with the highest Mn content.

\begin{table}
	\begin{tabular}{|c|c|c|c|c|c|c|}
		\hline
		Sample & Label & Mn $x_{\text{(av)}}$ &Thickness& Ga & Mn & Si \\
		\# & & (\%) &(nm) &  &  &  \\ \hline
		966 & (Ga,Mn)N & 0.5 & 470 & 5 & 490 & 0 \\
        1069 & (Ga,$\delta$Mn)N & 1.8 & 140 & 5 & $\delta$-490 & 0 \\
		1071 & (Ga,$\delta$Mn)N & 2.6 & 135 & 5 & $\delta$-490 & 0 \\
		1080 & (Ga,Mn)N & 1.1 & 740 & 5 & 490 & 0 \\
		1106 & (Ga,Mn)N & 1.8 & 750 & 1 & 490 & 0 \\
		1130 & (Ga,Mn)N& 1.8 & 200 & 1 & 490 & 0 \\
        1134 & (Ga,Mn)N:Si& 1.8& 220 & 1 & 490 & 2 \\
		1142 & (Ga,Mn)N & 3.1 & 230 & 1 & 490 & 0 \\
		1152 & (Ga,Mn)N:Si& 3.3 & 200 & 1 & 490 & 2 \\
		1159 & (Ga,$\delta$Mn)N & 1.5 & 140 & 1 & $\delta$-490 & 0 \\
		1160 & (Ga,$\delta$Mn)N:Si& 2.4 & 135 & 1 & $\delta$-490 & 2 \\
		1161 & (Ga,Mn)N:Si& 2.9 & 200 & 1 & 490 & 1 \\
		1268 & (Ga,Mn)N:Si& 2.0 & 232 & 1 & 400 & 2 \\
		1269 & (Ga,Mn)N:Si & 1.9 & 232 & 1 & 300 & 2 \\
		1273 & (Ga,Mn)N:Si& 0.49 & 780 & 5 & 490 & 2 \\
		1274 & (Ga,Mn)N & 0.53 & 780 & 5 & 490 & 0 \\
\hline

	\end{tabular}
	\caption{Samples studied in this work with Mn cation concentrations $x$ and $x_{\text{av}}$ for the uniformly [(Ga,Mn)N] and digitally Mn-doped structures [(Ga,$\delta$Mn)N], respectively, as determined by fitting the magnetic model to the data obtained by superconducting quantum interference device (SQUID), as described in Sec.~IV. The total thickness of the Mn-doped layer and precursor flow rates (in sccm) for the Ga, Mn, and Si containing precursors are also given. The samples \#1273 and 1274 have been grown on double-side epi-ready sapphire substrates suitable for optical transmission studies.}
\end{table}

\section{Nanocharacterization}
\subsection{Crystallinity}

The degree of crystallinity and possible precipitation of secondary crystallographic phases in uniformly and digitally Mn-doped films have been assessed by HR-TEM,  HR-STEM, HR-XRD, SXRD, and EXAFS.

Our TEM studies have been carried out for all studied films on cross-sectional samples prepared by mechanical polishing followed by Ar$^+$ ion milling, under a 4$^{\circ}$ angle at 4~kV for less than 2~h. The ion polishing has been performed in a Gatan 691 PIPS system.

The specimens have been investigated in Linz using a JEOL 2011 Fast TEM microscope operating at 200~kV and equipped with a Gatan CCD camera. The set-up is capable of an ultimate point-to-point resolution of 0.19~nm, with the possibility to image lattice fringes with a 0.14~nm resolution.

\begin{figure}[tb]
  \includegraphics[width=8cm]{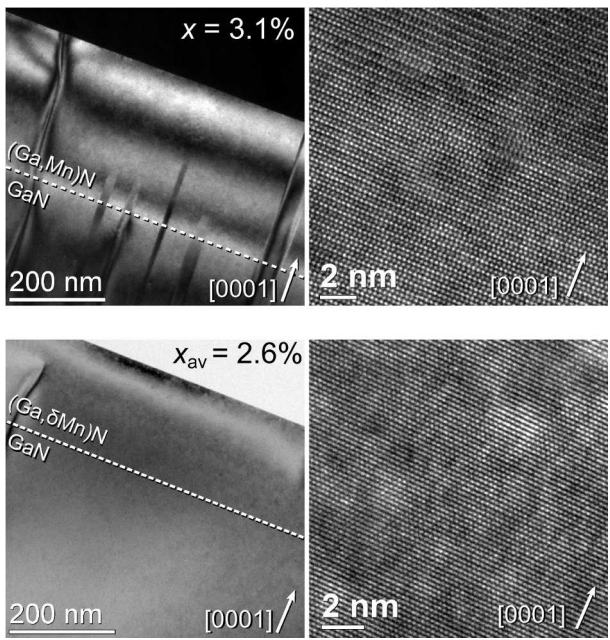}
  \caption{Low-magnification TEM (left panel) and HR-TEM (right panel) for (Ga,Mn)N ($x = 3.1$\%, upper panels) and (Ga,$\delta$Mn)N ($x_{\text{av}} = 2.6$\%, lower panels) without any evidence of crystallographic phase separation.}
  \label{fig:PRB_TEM}
\end{figure}

As reported in Fig.~\ref{fig:PRB_TEM}, for both (Ga,Mn)N ($x = 3.1$\%) and (Ga,$\delta$Mn)N ($x_{\text{av}} = 2.6$\%) samples low resolution TEM (left panels) shows no crystallographic phase separation and, in particular, no precipitates' segregation near the surface or interface. In fact, the HR-TEM images (right panels) clearly reveal the atomic positions in the lattice and, on the scale displayed, they show a homogenous crystal ordering and no signs of precipitation within the Mn-doped layers.

\begin{figure}[tb]
  \includegraphics[width=8cm]{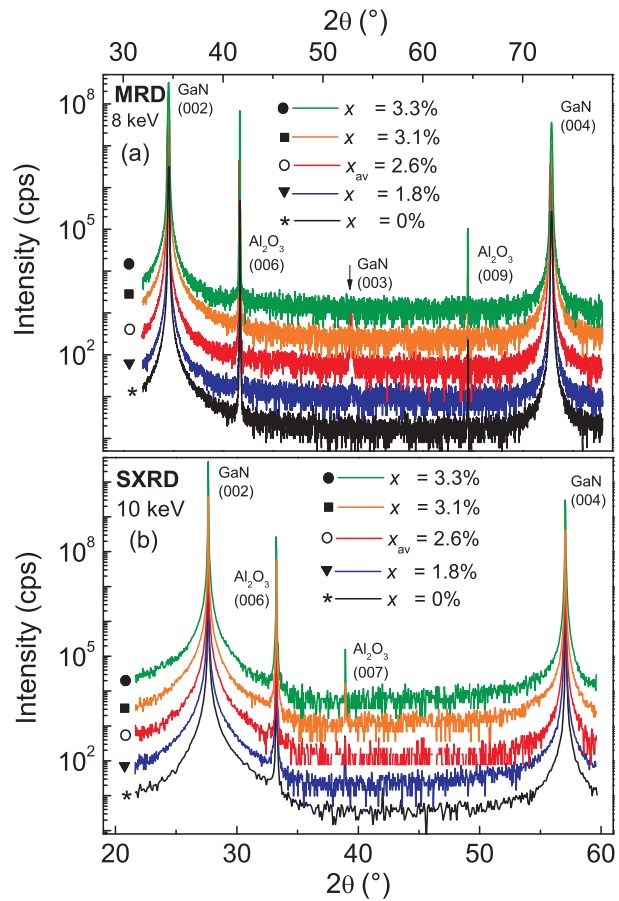}
\caption{(Color online) XRD radial scans of (Ga,Mn)N and (Ga,$\delta$Mn)N: (a) HR-XRD MRD, (b) SXRD collected at the BM20 of the ESRF.}
\label{fig:PRB_XRD}
\end{figure}

To further verify the crystallographic homogeneity of the grown Mn-doped layers, HR-XRD measurements using a Materials Research Diffractometer (MRD) and SXRD were performed. The HR-XRD experiments have been carried out with a Panalytical X'Pert PRO MRD in Linz at the photon energy of the Cu K$_{\alpha 1}$ radiation (8~keV) using a hybrid monochromator with a 0.25$^\circ$ slit for the incident optics and a pixel detector with an active length of 1~mm (19 channels) in the diffracted beam optics. The SXRD experiments have been performed at the beamline BM20 (Rossendorf Beam Line) of the European Synchrotron Radiation Facility (ESRF) in Grenoble using a point detector and 0.5 mm slits in front of the beam at an energy of 10~keV. Radial $\omega$-2$\theta$ scans of the GaN (002) to the GaN (004) diffraction peak do not show any trace of secondary phases. In Fig.~\ref{fig:PRB_XRD} the radial scans acquired with both techniques for the same set of samples are shown for comparison, with the MRD scans in the upper panel and the SXRD in the lower one. The sharp (002) and (004) diffraction peaks of GaN and the (006) diffraction of the sapphire substrate are observed for all samples. The measurements carried out with the high energy monochromatic beam at the synchrotron show a better signal-to-noise ratio and less diffuse scattering, when compared to those performed with the MRD, but in both cases no trace of crystallographic precipitation can be observed.

The narrow full-width at-half-maxima (FWHM) of the GaN (002) and (004) peaks (with values between 240 and 290 arcsec) indicate the high crystallinity of the layers. From the GaN symmetric sharp diffraction peak from the Mn-doped samples -- comparable to the one from the GaN reference -- we have hints that Mn-doping does not affect critically the dislocation density, as confirmed by the HR-TEM analysis on the same samples.

The XAFS measurements at the Mn K-edge (6539~eV) have been carried out at the GILDA Italian collaborating research group beam-line (BM08) of the ESRF in Grenoble, according to the experimental procedure detailed previously.\cite{Stefanowicz:2010_PRBa,Rovezzi:2009_PRB} Both EXAFS and XANES regions of the collected spectra have been analyzed.

\begin{figure}[htb]
  \centering
  \includegraphics[width=8cm]{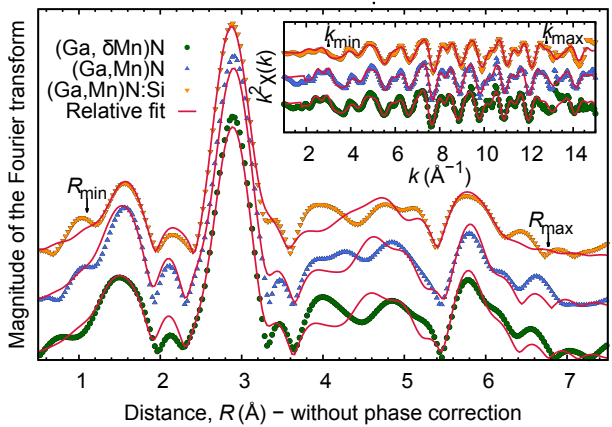}
  \caption{(Color online) Magnitude of the Fourier-transformed $k^{2}$-weighted EXAFS data (inset) in the range $k_{\mathrm{min}}$-$k_{\mathrm{max}}$ with relative fits (described in the text) in the range $R_{\mathrm{min}}$-$R_{\mathrm{max}}$ for representative samples: (Ga,$\delta$Mn)N ($x_{\mathrm{av}}$ = 2.6\%), (Ga,Mn)N ($x$ = 3.1\%) and (Ga,Mn)N:Si ($x$ = 3.3\%)}
  \label{fig:PRB_EXAFS}
\end{figure}

Three representative samples have been studied: (Ga,$\delta$Mn)N ($x_{\mathrm{av}}$ = 2.6\%), (Ga,Mn)N ($x$ = 3.1\%) and (Ga,Mn)N:Si ($x$ = 3.3\%). The collected data, with the polarization vector parallel to the $c$--axis, and the relative fits resulting from the EXAFS analysis are shown in Fig.~\ref{fig:PRB_EXAFS}. The structural model employed consists of one Mn atom substituting Ga in a GaN wurtzite crystal plus a Mn-Mn contribution taken from the MnN crystal structure\cite{Suzuki:2000_JAC} with a fitted coordination number (N$_{\mathrm{Mn-Mn}}$) in order to account for possible Mn clusters. The quantitative results are reported in Table~\ref{tab:EXAFS} and they are equivalent within the error bars for all the samples. The agreement with experimental data is good up to several coordination shells -- as seen in Fig.~\ref{fig:PRB_EXAFS} -- demonstrating the substitutional incorporation of Mn. The presence of an additional Mn-Mn coordination can be excluded within an uncertainty of 0.4 neighbors. Considering that the typical metal-metal coordination numbers are around 10 (12 Mn neighbors in MnN, 8 in Mn$_{3}$N$_{2}$) we can exclude the presence of secondary phases with a precision of about 5\%. In addition, the possible Mn incorporation in interstitial sites (tetrahedral and octahedral) has been investigated, statistically not improving the fit quality. These results are equivalent to the XAFS structural analysis on our (Ga,Mn)N samples at lower concentrations.\cite{Stefanowicz:2010_PRBa}

\begin{table}
  \caption{Quantitative results of the EXAFS analysis. For each sample, the fitted parameters are: the common amplitude (S$_{0}^{2}$), the average bond distances from the central Mn to the 4 N nearest neighbors ($R_{\mathrm{Mn-N}}$) and 12 Ga next nearest neighbors ($R_{\mathrm{Mn-Ga}}$) of the wurtzite structure plus a common expansion parameter for higher coordination shells ($\Delta$$R$) and the coordination number (N$_{\mathrm{Mn-Mn}}$) for the Mn-Mn bond distance at 2.98(2) \AA. The Debye-Waller parameters attest all below 8(2)\,$\times$\,10$^{\mathrm{-3}}$ \AA{}$^{\mathrm{-2}}$ and a correlated Debye model\cite{Sevillano:1979_PRB} with a temperature of 500(50) K is used for the GaN multiple scattering contributions. Error bars are reported on the last digit.}
  \begin{center}
    \begin{tabular}{|l|c|c|c|c|c|c|}
      \hline
      Sample & $x_{\text{(av)}}$& S$_{0}^{2}$ & $R_{\mathrm{Mn-N}}$ & $R_{\mathrm{Mn-Ga}}$ & $\Delta$$R$ & N$_{\mathrm{Mn-Mn}}$ \\
      &  & (\%)& (\AA)  &  (\AA)  & (\%) &  \\
      \hline
      (Ga,$\delta$Mn)N & 2.6& 0.90(5)  & 1.95(3)  & 3.20(2)  & 0.1(2)  & 0.4(4)  \\
      (Ga,Mn)N       &  3.1& 0.94(5)  & 1.94(3)  & 3.19(2)  & 0.1(2)  & 0.2(4)  \\
      (Ga,Mn)N:Si &3.3     & 0.94(5)  & 1.96(3)  & 3.19(2)  & 0.1(2)  & 0.0(4)  \\
      \hline
    \end{tabular}
    \label{tab:EXAFS}
  \end{center}
\end{table}

In conclusion, TEM, HR-(S)TEM, HR-XRD, SXRD, and EXAFS studies demonstrate a single-crystal character of the considered films with no traces, down to the atomic scale, of precipitations (crystallographic phase separation). This is in contrast with the case of some (Ga,Mn)N films grown by MBE (Refs.~\onlinecite{Dhar:2003_APL,Giraud:2004_EPL}) as well as with (Ga,Fe)N layers obtained by MOVPE (Refs.~\onlinecite{Bonanni:2007_PRB,Bonanni:2008_PRL,Rovezzi:2009_PRB,Navarro-Quezada:2010_PRB}), for which crystallographic phase separations were detected by employing the same nanocharacterization tools.

\subsection{Mn distribution}

\begin{figure}[htbp]
  \centering
  \includegraphics[width=8.5cm]{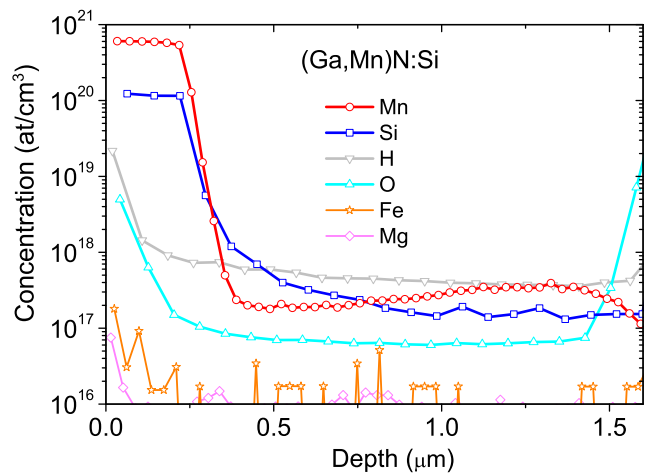}
  \caption{(Color online) SIMS depth profiles of Mn and Si in the (Ga,Mn)N:Si film with $x = 1.9$\%. The residual contamination by H, O, Fe, and Mg is also shown.}
  \label{fig:PRB_SIMS}
\end{figure}

The Mn distribution along the growth direction and the Si distribution in the co-doped films have been evaluated in Warsaw via SIMS, calibrated by Mn implanted GaN, providing the absolute concentration of Mn atoms with an accuracy of about a factor 2. The SIMS depth profiles reported previously\cite{Stefanowicz:2010_PRBa} for the samples with $x \lesssim 1$\% and the one shown in Fig.~\ref{fig:PRB_SIMS} for the present films indicate that the Mn distribution  is  uniform over the Mn-doped region and that the interface between the (Ga,Mn)N overlayer and the GaN buffer layer is sharp.

This is confirmed by chemical analysis performed in Linz with an Oxford Inca EDS system, which---with the sensitivity of 0.1\% at.---does not provide any evidence for Mn diffusion into the nominally undoped GaN buffer.

\begin{figure}[tb]
\includegraphics[width=8.5cm]{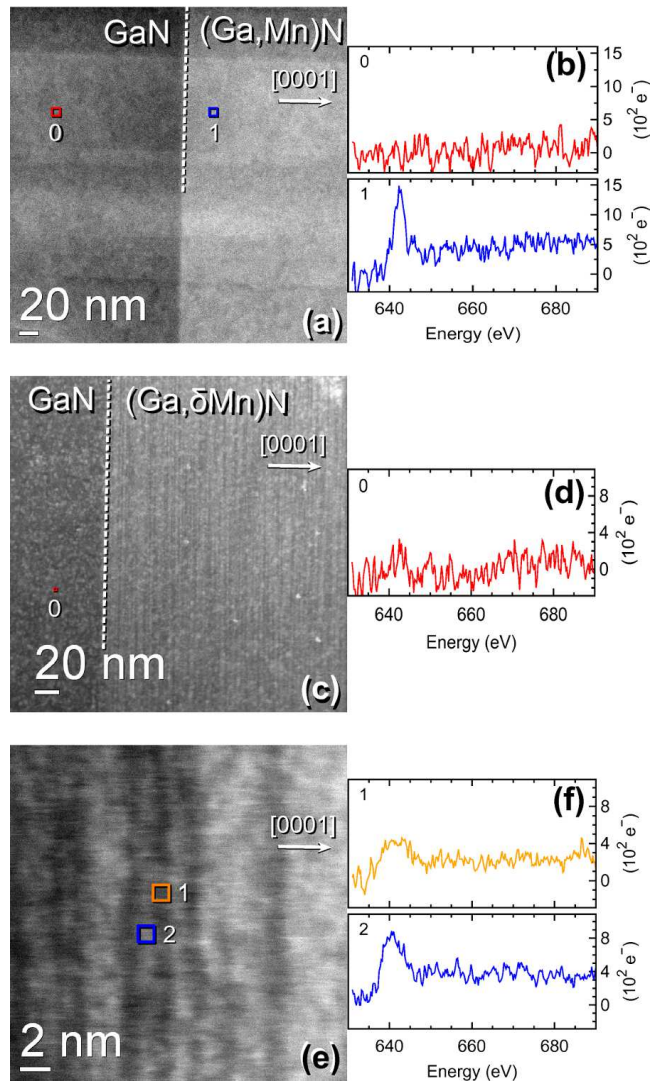}
\caption{(Color online) (a) HAADF-STEM image of (Ga,Mn)N ($x = 3.1$\%)-- change in intensity (chemical contrast), when going from the substrate into the doped layer (from left to right).  (b) EELS spectra for points 0 (GaN buffer layer) and 1 (nominally Mn doped layer): evidence of the presence of Mn only in the Mn-doped layer (right side of the image). (c-f) Determination of the Mn distribution in the digitally Mn-doped sample (Ga,$\delta$Mn)N ($x_{\text{av}} = 2.6$\%, 160 Mn periods, layer thickness 135~nm). (c,e) HAADF-STEM scans in the Mn-doped region giving modulated chemical contrast. (d,f)  EELS spectra of points 0, 1, and 2, as marked in (c) and (e), respectively.}
\label{fig:PRB_TEM_EELS}
\end{figure}

For further characterization of the Mn doping, HAADF-STEM and EELS measurements have been performed -- and reported in Fig.~\ref{fig:PRB_TEM_EELS} -- employing a FEI Tecnai F 20 200kV transmission electron microscope in Graz. The STEM images and EELS spectra could be recorded with an upgraded spectrometer with an adapted STEM detection geometry, optimized for $Z$-contrast and enhanced spectral collection EELS efficiency.\cite{Gubbens:2010_UM,Riegler:2010_UM} In order to improve the signal to noise ratio at the high spatial resolution in question as well as to keep exposure times low for minimal sample drift, the EELS point spectra have been taken as an integral sum over the energy. Accordingly, a possible fine structure located within the first 20--30~eV around the ionization threshold energy of an edge has been averaged out.

The HAADF-STEM on (Ga,Mn)N ($x = 3.1$\%) and reported in Fig.~\ref{fig:PRB_TEM_EELS}(a) reveals a clear change in intensity, when going from the substrate into the Mn-doped layer (from left to right), while within the doped layer no chemical contrast could be detected. The EELS spectra given in Fig.~\ref{fig:PRB_TEM_EELS}(b) evidence the presence of Mn only in the nominally Mn-doped layer.

In contrast, remarkably, as shown in Figs.~\ref{fig:PRB_TEM_EELS}(c) and \ref{fig:PRB_TEM_EELS}(e), HAADF-STEM observations on the digital (Ga,$\delta$Mn)N ($x_{\mathrm{av}}$ = 2.6\%) reveal intensity modulations. Additional EELS point spectra collected in Fig.~\ref{fig:PRB_TEM_EELS}(d) show the signal differences of the Mn L$_{23}$ edge between the substrate and the intensity modulated lines.

In conclusion, the element specific analysis demonstrates a spatially homogeneous Mn distribution over the volume of the uniformly Mn-doped (Ga,Mn)N films, with no segregation towards the surface, interface or buffer regions. In contrast, in the case of the digitally  Mn-doped films, nano-scale density modulation with a period imposed by the growth conditions has been detected, meaning that in these $\delta$Mn samples the local Mn concentration fluctuates between lower and higher $x$ values around the $x_{\text{av}}$ determined by SQUID magnetometry. As discussed in Sec.~IV, such a non-random distribution of Mn ions increases the apparent Curie constant, particularly if the system is close to a ferromagnetic instability.

\subsection{Concentrations of {Si} and {Mn$^{2+}$} ions}

The incorporation and a uniform distribution of Si impurities in (Ga,Mn)N:Si layers is evidenced by the SIMS result displayed in Fig.~\ref{fig:PRB_SIMS}. From the same measurements on all the considered samples, the Si concentration is found to be of the order of $10^{20}$~cm$^{-3}$ for the Si and Ga precursor flow rates 1 or 2 and 1~sccm, respectively.

Following recent works of the Mn charge state in GaN,\cite{Wolos:2008_B} the effect of co-doping by Si donors on the Mn charge state has been quantitatively assessed by examining the magnitude of the intra-ion optical absorption, which occurs at $E_0 \approx 1.4$~eV for Mn$^{3+}$ ions in GaN.\cite{Korotkov:2001_PB,Wolos:2004_PRBb,Zenneck:2007_JAP,Malguth:2008_MRS} Optical investigations have been performed in Warsaw and in Linz for two series of films, abridged in Table~\ref{table:fits}, with the Mn concentrations $x \approx 1.8$ and 0.5\%, respectively, and different Si content. The series with the lower Mn concentration has been designed for transmission and ESR measurements with the layers deposited on double-side epi-ready sapphire substrates.

\begin{figure}
  \includegraphics[width=8.3cm]{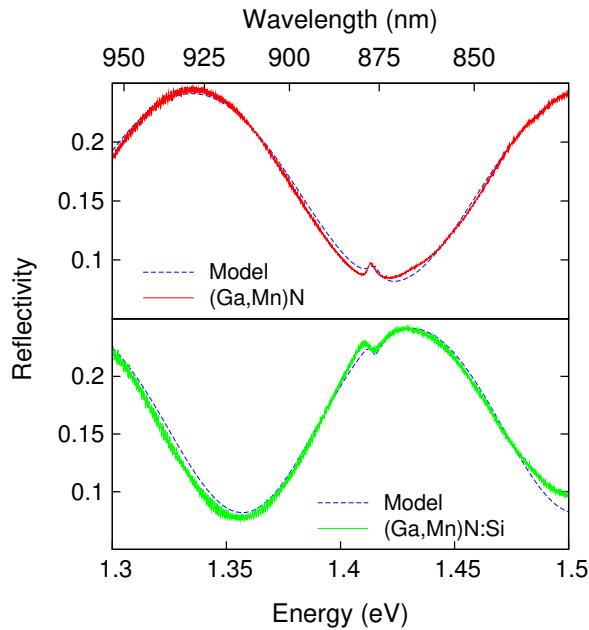}
  \caption{(Color online) Optical reflectivity spectra for (Ga,Mn)N without and with Si co-doping (upper and lower panel, respectively).  The Mn concentration is 1.8\% in both samples. The absorption feature at 1.41~eV is clearly resolved. The fitting results by the transfer matrix multilayer model are given by the dashed lines.}
  \label{fig:PRB_reflectivity}
\end{figure}

\begin{figure}
  \includegraphics[width=8.3cm]{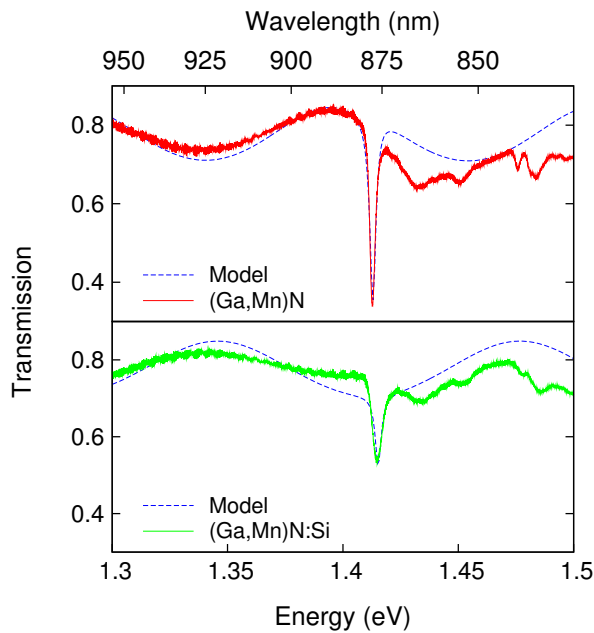}
  \caption{(Color online)  Optical transmission spectra for samples with a Mn concentration $x \approx 0.5$\%. Upper and lower graphs: data for samples without and with Si co-doping, respectively.  Next to the main absorption line at 1.41~eV the phonon replicas of the line are observed. The fitting results by the multilayer model are given by the dashed lines.}
  \label{fig:PRB_transmission}
\end{figure}

As it can be seen in  Figs.~\ref{fig:PRB_reflectivity} and \ref{fig:PRB_transmission}, the Si doping quenches the intra-ion absorption, specific to Ga-substitutional Mn in the 3+ charge state.

We determine the concentration ratio of the absorbing $\mathrm{Mn^{3+}}$ ions for samples without and with Si by fitting a model constructed within the transfer matrix formalism of optical transmission and reflectivity, taking into account the doped (Ga,Mn)N layer, the undoped GaN buffer, and the sapphire substrate.\cite{Katsidis:2002_AO}


The contribution of the Mn$^{3+}$ ions to the dynamic dielectric function is modeled by damped Lorentzian oscillators of the form,
\begin{equation}
  \epsilon(\omega)=\epsilon_{\mathrm{GaN}}(\omega) + \frac{ f N_{\text{Mn3+}}}{
E_0^2-(\hbar\omega)^2-i \hbar\omega \Gamma}
  \label{eqn:lorentz}
\end{equation}
Here $f$ is proportional to the oscillator strength and  $\Gamma$ is the damping energy. The refractive index of GaN is modeled using the Sellmeier equation\cite{Yu:1997_APL} and the refractive index of the sapphire is set at 1.8.  The  thicknesses of the GaN buffer and (Ga,Mn)N layer are adjusted to reproduce the observed Fabry-P\'erot interferences at a magnitude of thicknesses ratio as determined during the growth by \textit{in-situ} ellipsometry.

In Table~\ref{table:fits} the fitted values of the parameters in Eq.~1 are given. From the reduction of the Mn$^{3+}$ absorption we can evaluate the concentration of Mn$^{2+}$ ions assuming that electrons coming from Si shallow donors occupy the Mn$^{2+}$/Mn$^{3+}$ midgap level. Under this assumption, the concentration of Mn$^{2+}$ ions reaches a level of 4$\times$10$^{20}$ per cm$^{3}$ for the highest employed flow of the Si precursor at the lowest Ga precursor flow.

\begin{table}[t,b]
  \begin{center}
    \begin{tabular}{|c|c|c|c|c|c|}
      \hline
      Label & $x$ & $E_0$    & $f N_{\mathrm{Mn^{3+}}}$ & $\Gamma$  & Mn$^{\text{2+}}$ \tabularnewline
      &(\%)    & (meV)      & (meV$^2$)            &  (meV)   &($10^{20}/\mathrm{cm}^3$)\tabularnewline
      \hline
      (Ga,Mn)N &1.8 & 1415& 13000 & 6.1& \tabularnewline
      (Ga,Mn)N:Si&1.8& 1414   & 5300& 3.6 &4$\pm$1\tabularnewline
       \hline
      (Ga,Mn)N &0.53 & 1413 & 950& 2.57& \tabularnewline
      (Ga,Mn)N:Si & 0.49 & 1415   & 490& 3.0  & 0.6$\pm$0.1  \tabularnewline
    \hline
    \end{tabular}
  \end{center}
 \caption{Values of parameters in Eq.~1 determined from reflectivity and transmission measurements for two series of samples (the upper and lower panel, respectively).  The concentration of Mn$^{2+}$ ions is calculated according to $x^{\mathrm{(Ga,Mn)N:Si}}-x^{\mathrm{(Ga,Mn)N}}r$, where $r$ is the ratio of the $f N_{\mathrm{Mn^{3+}}}$-values within a given series of samples. }
 \label{table:fits}
  \end{table}

\begin{figure}
 \includegraphics[width=8.3cm]{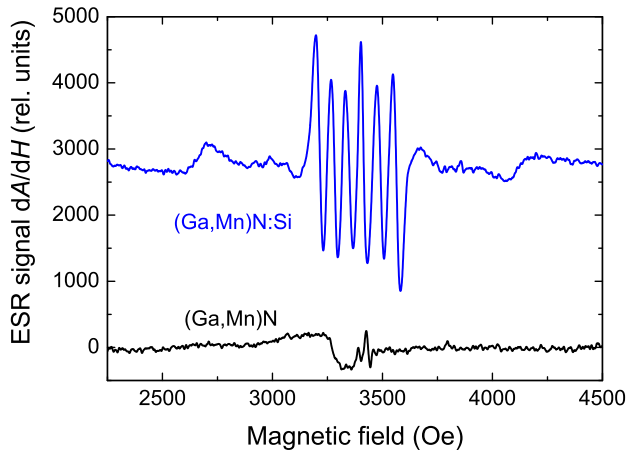}
\caption{(Color online) Results of ESR studies at 2~K for (Ga,Mn)N without and with Si co-doping (lower and upper panel, respectively) showing the emergence of  Mn$^{2+}$ signal on Si doping. The Mn concentrations is $x \approx 0.5$\%.}
\label{fig:PRB_ESR}
\end{figure}

Results of ESR studies carried out in Linz and reported in Fig.~\ref{fig:PRB_ESR} for (Ga,Mn)N and (Ga,Mn)N:Si with the Mn concentration $x \approx 0.5$\% demonstrate the emergence of a characteristic Mn$^{2+}$ signal upon Si doping.  In films with higher Mn concentrations, the line broadening, witnessing the presence of Mn$^{2+}$--Mn$^{3+}$ coupling, as discussed in Sec.~IV, has hampered the detection of a Mn$^{2+}$ signal.

A non-zero value of the orbital momentum, and the associated spin-orbit interaction specific to Mn$^{3+}$ ions, precludes their observation by ESR (Ref.~\onlinecite{Wolos:2008_B}). At the same time, Mn$^{2+}$ ions, corresponding to orbital singlets, give rise to a specific ESR response.\cite{Wolos:2008_B,Graf:2003_PRB}

\begin{figure}[htb]
  \centering
  \includegraphics[angle=-90,width=8.3cm]{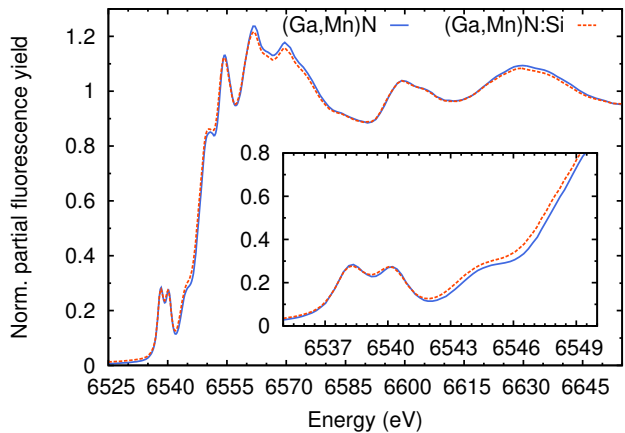}
  \caption{(Color online)  Partial fluorescence yield XANES spectra (integrated K$_{\alpha 1,2}$ fluorescence lines) for (Ga,Mn)N and (Ga,Mn)N:Si samples with the polarization vector parallel to the wurtzite $c$ axis. Inset: focus on the shift of the main absorption edge between the two spectra. The position of the pre-edge peaks is constant.}
  \label{fig:PRB_XANES}
\end{figure}

The results of the optical and ESR studies are confirmed also by XANES. In order to contribute to the determination of the Mn valence state, the position of the x-ray absorption edge and the pre-edge features have been considered as described in the XANES section of Ref.~\onlinecite{Stefanowicz:2010_PRBa}. In particular here the issue of reduction of the charge state is addressed. As it can be appreciated in Fig.~\ref{fig:PRB_XANES}, a shift towards lower energies is visible for the Si co-doped sample, while the position of the pre-edge peaks remains unchanged. This demonstrates the fine calibration of the incoming energy and suggests the presence of some Mn in a valence state lower than 3+, possibly 2+. Due to the lack of model compounds that would allow to establish a precise relation between the edge shift and the valence state in nitrides, a quantitative statement cannot be given here. However, considering that in the case of 6-coordinated Mn ions (this example is taken as no data are reported for tetragonal Mn$^{\mathrm{3+}}$) the ionic radius of Mn$^{\mathrm{2+}}$ is about 12\% greater than the one of Mn$^{\mathrm{3+}}$ (Ref.~\onlinecite{Shannon:1976_ACA}) and that no visible evolution of the Mn-N distance is reported, we can expect  a minority of Mn ions to be in the 2+ charge state.

In conclusion, our SIMS, optical, XANES, and ESR studies show consistently that co-doping with Si increases the concentration of Mn ions in the 2+ charge state, which for the highest employed flow of the Si precursor (2 sccm) and the lowest Ga precursor flow is up to 4$\times$10$^{20}$~cm$^{-3}$, about 30\% of the total Mn concentration for $x = 3$\%. This evaluation substantiates the experimental data presented in the next section. It is worth noting that a co-existence of Mn$^{3+}$ and Mn$^{2+}$ ions was detected also in x-ray magnetic circular dichroism in (Ga,Mn)N samples undoped with Si,\cite{Freeman:2007_PRB} pointing to the presence of residual or interfacial compensating donors.

\section{Magnetic properties}

\begin{figure}[tb]
  \centering
   \includegraphics[width=8.5cm]{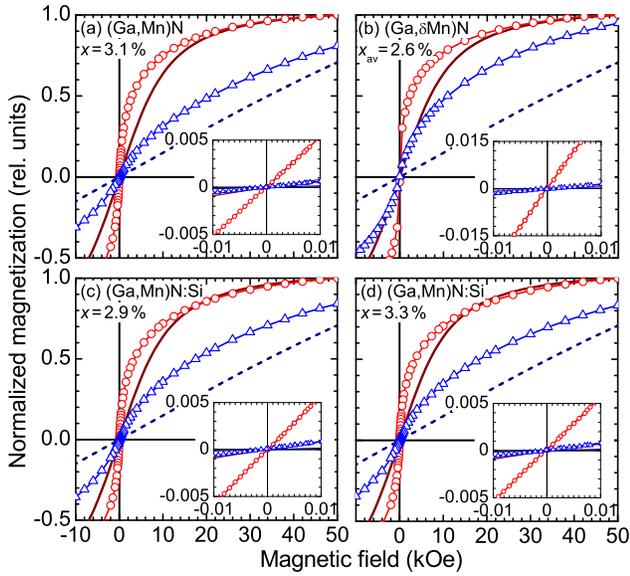}
  \caption{(Color online) Magnetization at 1.85~K (normalized to its in-plane value at 50~kOe) as a function of the magnetic field applied perpendicular (circles) and parallel (triangles) to the wz $c$-axis of (Ga,Mn)N (a,b) and (Ga,Mn)N:Si (c,d) films where Mn is introduced either uniformly (a,c,d) or digitally (b). Solid and dashed lines are calculated  according to the group theoretical model for non-interacting Mn$^{3+}$ ions in wz-GaN for $H$ perpendicular and parallel to the $c$-axis, respectively.\cite{Stefanowicz:2010_PRBa} Insets: low-field magnetization loops.}
  \label{fig:PRB_M_H}
\end{figure}

\begin{figure}[tb]
  \centering
\includegraphics[width=9cm]{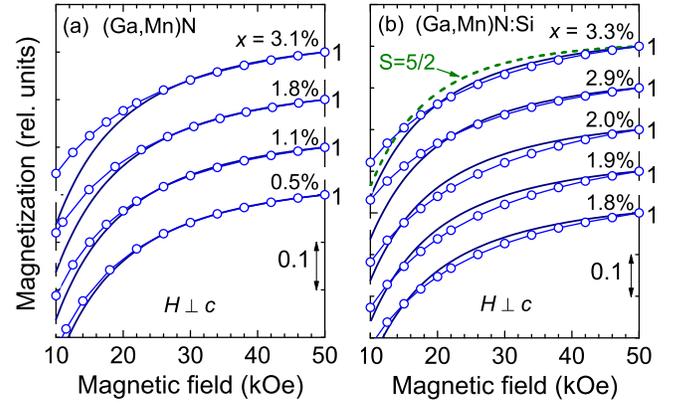}
 \caption{Comparison of magnetization saturation as a function of the magnetic field between (a): (Ga,Mn)N -- no antiferromagnetic interactions and (b): (Ga,Mn)N:Si -- with antiferromagnetic interactions. The relative experimental error is about one half of the point size. Solid lines and the dashed line in (b) are calculated for non-interacting Mn$^{3+}$ and Mn$^{2+}$ ions, respectively.}
  \label{fig:PRB_M_H_Si}
\end{figure}

According to our previous studies of (Ga,Mn)N with $x < 1$\%,\cite{Stefanowicz:2010_PRBa} the dependence of the magnetization $M$ on temperature $T$, magnetic field $H$, and its orientation with respect to the wurtzite (wz) $c$-axis can accurately be described in terms of non-interacting Mn$^{3+}$ ions substitutional of Ga. The good agreement between the experimental data and the model confirms a weak compensation by residual impurities which, if present, would change the Mn charge state and thus the magnetic properties. The Mn$^{3+}$ charge state is preserved in samples with higher Mn concentrations, where the persistence of a large anisotropy between the $M(H)$ values at 1.85~K for the two sample orientations $c \perp H$ and $c \parallel H$ is evidenced in Figs.~\ref{fig:PRB_M_H}(a) and \ref{fig:PRB_M_H}(b). However, a gradual enhancement of $M(H)/M(50\,\mbox{kOe})$ over the magnitude expected for non-interacting spins is observed when increasing $x$ up to 3\% in both uniformly and digitally Mn-doped films, as seen in Figs.~\ref{fig:PRB_M_H}(a), \ref{fig:PRB_M_H}(b), and \ref{fig:PRB_M_H_Si}(a). These results demonstrate univocally that, in spite of the absence of band carriers, the dominant exchange interaction between Mn$^{3+}$ is {\em ferromagnetic} in (Ga,Mn)N.

Interestingly, a rather different behavior is observed in the case of (Ga,Mn)N:Si, where the trapping of donor electrons changes the Mn charge from 3+ to 2+ and the spin state $S$ from 2 to 5/2, for about 30\% of the Mn ions at $x \approx 3$\%, as discussed in Sec.~IIIC.
According to the data collected in Figs.~\ref{fig:PRB_M_H}(c), \ref{fig:PRB_M_H}(d), and \ref{fig:PRB_M_H_Si}(b) the increasing concentration of Mn$^{2+}$ ions results in the foreseen decrease of the magnetization anisotropy. Furthermore, as shown in Fig.~\ref{fig:PRB_M_H_Si}(b), $M(H)$ saturates {\em slower} than theoretically anticipated for non-interacting Mn ions. This finding points to an {\em antiferromagnetic} character of the exchange coupling between Mn$^{2+}$ ions, and suggests that these ions may dominate in (Ga,Mn)N, when no ferromagnetic interactions are detected.\cite{Zajac:2001_APL,Granville:2010_PRB}

In the insets to Figs.~\ref{fig:PRB_M_H}(a)--\ref{fig:PRB_M_H}(d) the results of our search for the onset of a collective magnetic behavior in the samples with the highest Mn concentrations are given. A linear and ahysteretic $M(H)$ dependence in weak magnetic fields is observed for both configurations, $c \bot  H$ and $c \| H$, pointing to the absence of spontaneous magnetization.  These data imply that the ferromagnetic spin-spin couplings are too short--ranged to result in magnetic ordering and, hence, in spontaneous magnetization at $T \ge 1.85$~K.

\begin{figure}[tb]
  \centering
  \includegraphics[width=9.5cm]{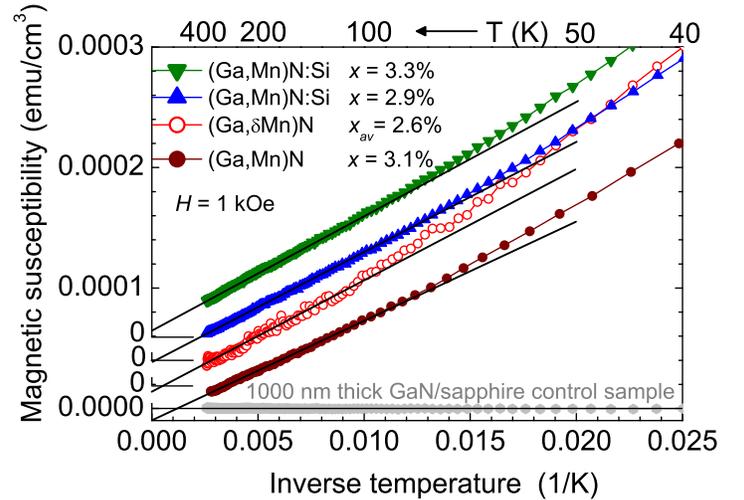}
  \caption{(Color online) Magnetic susceptibility $M/H$ for $c\perp H$ as a function of the inverse temperature for uniformly (solid symbols) and digitally (open symbols) doped (Ga,Mn)N (circles) and (Ga,Mn)N:Si (triangles) as well as GaN/sapphire control sample (light grey circles). Solid lines serve to determine the Curie constants at high temperatures. Deviations of their values form zero at $1/T = 0$ show the accuracy of the substrate subtraction.}
  \label{fig:PRB_chi_T}
\end{figure}

Quantitative information on $J_{\text{nn}}$ is gained here by examining the dependence  on the inverse temperature of the in-plane magnetic moment $m(T)$ of Mn spins in GaN, as obtained by subtracting the value of $m(T)$ measured independently for a sapphire substrate (normalized by the corresponding sample weight). As reported in Fig.~\ref{fig:PRB_chi_T}, $\chi(T) \equiv M/H \sim 1/T$ for $150 \lesssim T \leqslant 350$~K.  This behavior indicates that the contribution to the magnetic susceptibility from the (Ga,Mn)N films obeys the Curie law in this regime, $\chi(T)= C/T$. This dependence is expected if the spin pairs are either uncorrelated $|J_{\text{nn}}|S^2 \ll k_{\text{B}}T$, or strongly bound $|J_{\text{nn}}|S^2 \gg k_{\text{B}}T$.

In order to extract from these data the magnitude of $J_{\text{nn}}$, we extend the previous model of a single substitutional Mn$^{3+}$ impurity in GaN (Ref.~\onlinecite{Stefanowicz:2010_PRBa}) by considering a pair of Mn$^{3+}$ ions coupled by an exchange interaction  $-J\vec{S}_1\vec{S}_2$,\cite{Spalek:1986_PRB,Shapira:2002_JAP} the model discussed in details in the Appendix. Assuming a random distribution of Mn over the cation hcp lattice, and allowing for the coupling between nn spins we can evaluate $M(T,H)$ at a given $J_{\text{nn}}$ and $x$. This approach implies, in particular, that for $x =3$\%, $T < 350$~K, and $H = 1$~kOe, $\chi(T) = C/T$ if $J_{\text{nn}} > 10$~meV. However, in this case, due to the proportionality of $\chi(T)$ to the pair spin square, the magnitude of $C$ is enhanced in comparison to the value $C_0$ corresponding to non-interacting spins. To evaluate experimentally $C_{\text{norm}} = C/C_0$, we consider that its magnitude can be determined from the magnetic moment $m(T,1\,\mbox{kOe})$ measured in-plane without knowing the exact value of the volume occupied by the Mn spins, if the magnitude of the in-plane $m(1.85\,\mbox{K},50\,\mbox{kOe})$ is employed to obtain the Mn content $x$ --- and thus $C_0$ --- for particular samples. Following the outcome of the experimental results for (Ga,Mn)N:Si (Sec.~IIIC), demonstrating the presence of Mn$^{2+}$ ions, their relative contribution to $M(T,H)$ is determined from the magnitude of the magnetic anisotropy.

\begin{figure}[tb]
  \centering
    \includegraphics[width=8.5cm]{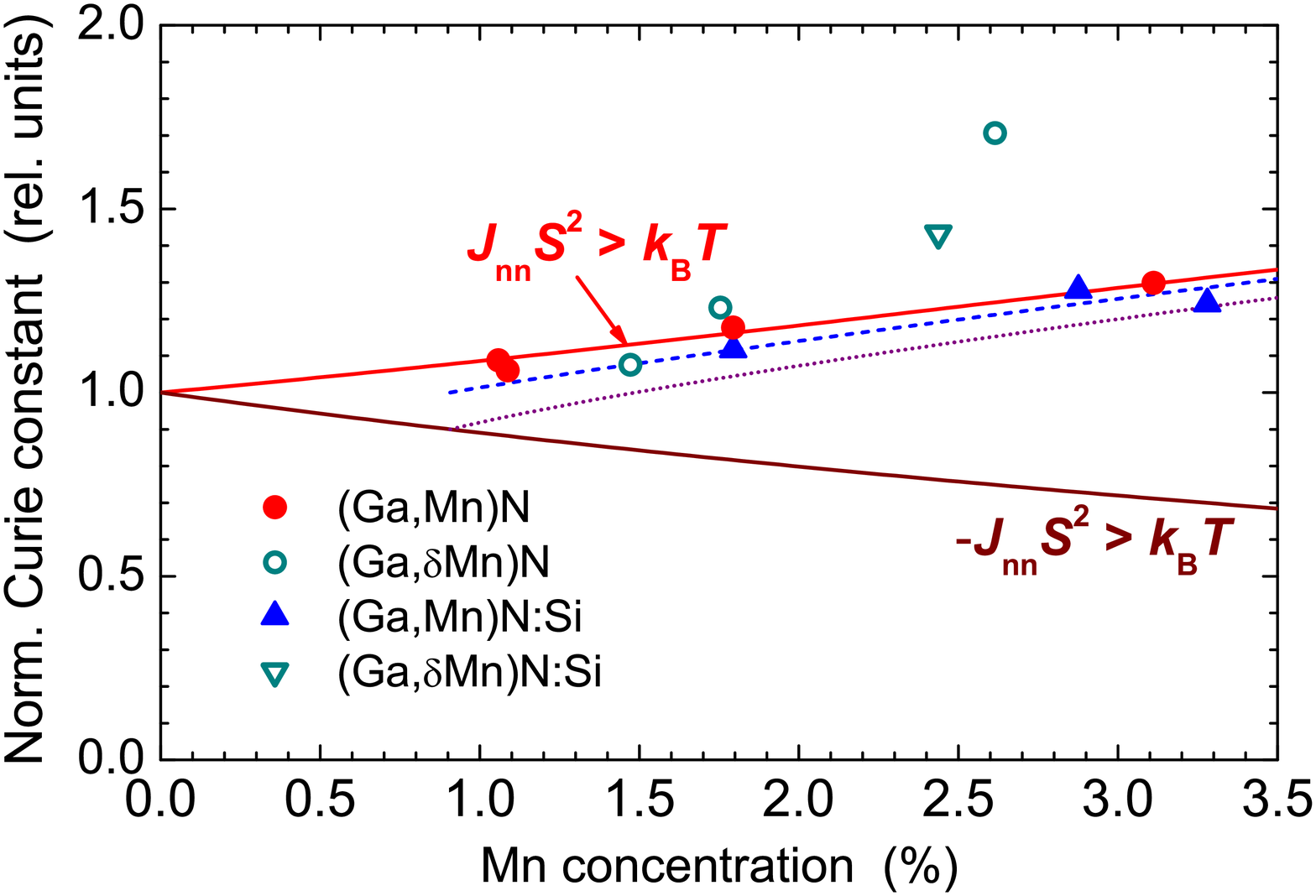}
  \caption{(Color online)  Normalized Curie constant $C_{\text{norm}}$ as a function of the Mn content for uniformly (solid symbols) and digitally (open symbols) doped (Ga,Mn)N (circles) and (Ga,Mn)N:Si (triangles). The solid lines are computed assuming a random distribution of Mn$^{3+}$ ions, and either ferromagnetic or antiferromagnetic strong coupling between the nearest neighbor (nn) Mn spins, $|J_{\text{nn}}|S^2 \gg k_{\text{B}}T$. Dashed and dotted lines are calculated assuming that $4\cdot 10^{20}$~cm$^{-3}$ Mn ions are in the $2+$ charge state, and nn interactions are ferromagnetic except for nn Mn$^{2+}$ pairs for which $J_{\text{nn}} = 0$ or $-J_{\text{nn}}S^2 \gg k_{\text{B}}T$, respectively.}
  \label{fig:PRB_C_x}
\end{figure}

As summarized in Fig.~\ref{fig:PRB_C_x}, $C_{\text{norm}} >1$ for all studied samples, hinting to the presence of a considerable ferromagnetic spin-spin interaction. The theory presented in the Appendix describes quite well the magnitude of $C_{\text{norm}}(x)$ for the uniformly doped (Ga,Mn)N films, pointing to $J_{\text{nn}} > 10$~meV, in general agreement with the results of the {\em ab initio} studies outlined in Fig.~\ref{fig:PRB_JNN}. Furthermore, a low-temperature upturn of the experimental points over the $C/T$ dependence, visible in Fig.~\ref{fig:PRB_chi_T} below $\sim$100~K, suggests the existence of an additional weak ferromagnetic coupling between more distant neighbors. Moreover, the experimental values of $C_{\text{norm}}(x)$ in the case of digital $\delta$Mn-doping are higher than theoretically expected.

In order to clarify the different magnitude of  $C_{\text{norm}}$ in uniformly and digitally doped films, we refer to Sec.~IIIB, where the detailed investigation of the Mn distribution for the two samples with the highest Mn concentration, respectively (Ga,Mn)N ($x = 3.1$\%) and the digital (Ga,$\delta$Mn)N ($x_{\mathrm{av}}$ = 2.6\%), have been shown. The data demonstrate the presence of a spatially modulated Mn concentration in the digitally Mn-doped films. Due to a non-linear dependence of the Curie constant on the Mn concentration in the presence of ferromagnetic interactions, such a non-random distribution of Mn ions increases the apparent value of $C_{\text{norm}}$, particularly if the system is close to a ferromagnetic instability. This interpretation is supported by a much smaller effect in the films with lower values of $x_{\text{av}}$, and thus far from the ferromagnetic instability.

Finally, we comment on the magnitudes of $C_{\text{norm}}(x)$ in Si doped samples. Here, we have ferromagnetically coupled Mn$^{3+}$--Mn$^{3+}$ and Mn$^{3+}$--Mn$^{2+}$ pairs as well as antiferromagnetically interacting Mn$^{2+}$--Mn$^{2+}$ pairs. As shown in Fig.~\ref{fig:PRB_C_x}, the theory developed for such a case and summarized in the Appendix is consistent with the data for Si-doped samples.

\section{Conclusions}

In this work, we have verified experimentally the presence of a strong ferromagnetic coupling between neighboring Mn spins in (Ga,Mn)N, supporting a very significant number of {\em ab initio} studies.\cite{Sato:2001_JJAP,Sato:2010_RMP,Zunger:2010_P} Since the Mott-Hubbard localization precludes carrier hopping between magnetic ions, a ferromagnetic super-exchange constitutes the relevant microscopic coupling mechanism.\cite{Blinowski:1996_PRB} However, according to our findings, the range of this interaction is too short to produce a long-range ferromagnetic ordering -- at least above 1.85~K -- in samples with 3\% of randomly distributed substitutional Mn cations. Co-doping with Si may {\em a priori} result in a ferromagnetic double exchange, but apparently Anderson-Mott localization in the Mn impurity band renders this mechanism inefficient in the range of Mn contents explored so far by us. If, owing to a large density of donor-like defects or impurities the concentration of Mn$^{2+}$ prevails, antiferromagnetic super-exchange becomes the dominant spin-spin coupling mechanism. This situation has presumably taken place in recently studied  (Ga,Mn)N films with $x$ up to $36$\% (Ref.~\onlinecite{Granville:2010_PRB}) and a time ago in the case of In$_{1-x}$Mn$_x$As layers with $x$ up to $18$\%.\cite{Munekata:1989_PRL}

\begin{acknowledgements}

The work was supported by FunDMS Advanced Grant of ERC (Grant No. 227690) within the Ideas 7th Framework Programme of European Community, InTechFun (Grant No.
POIG.01.03.01-00-159/08), SemiSpinNet (Grant No. PITNGA-
2008-215368), by the Austrian FWF (P20065, P22477, P20550) and FFG (N107-NAN), and by the NCBiR project LIDER.

\end{acknowledgements}

\section*{Appendix: Theoretical evaluation of the Curie constant}

We evaluate the Curie constant $C_{\text{norm}}$ for a random distribution of the Mn$^{3+}$ and Mn$^{2+}$ ions. While our model can be applied for a general situation, we discuss the case describing our data in the relevant temperature range $150 \lesssim T \leqslant 350$~K,  {\em i.e.}, we assume that the nearest neighbour (nn) Mn$^{3+}$--Mn$^{3+}$ and Mn$^{3+}$--Mn$^{2+}$ pairs form ferromagnetically oriented dimers, whereas nn Mn$^{2+}$--Mn$^{2+}$ pairs are uncoupled.

In the paramagnetic region well above the ordering temperature, (high temperature limit) the magnetic susceptibility  is expected to obey the Curie-Weiss law,
\begin{eqnarray}
\label{eq:CurieWeiss}
\chi=\frac{C_0}{T-\theta_C} \\
C_0=N\frac{{(g\mu_{\text{B}})}^2S(S+1)}{3k_{\text{B}}}
\end{eqnarray}
where $C$ and $\theta_C$ are the Curie constant and Curie-Weiss temperature, and $N$ is the concentration of magnetic ions with spin $S$. However, in random magnetic alloys, where the interactions between spin pairs show a large dispersion owing to strong variations of the spin-spin distances, the magnitudes of $C$ and $\theta_C$ may depend on the temperature.\cite{Spalek:1986_PRB,Shapira:2002_JAP,Bhatt:1982_PRL}

We consider the case of dilute magnetic semiconductors (DMSs) and dilute magnetic oxides (DMOs). In the absence of band carriers that could mediate long range spin-spin interactions, the strength of the exchange couplings decays rather fast with the spin-spin distance. In such a case, the exchange between magnetic ions occupying the nn cation positions dominates. Accordingly, below the percolation limit for the nn interaction ($x \lesssim 18$\% for fcc and hcp lattices), the magnetic response can be evaluated as a sum of contributions coming from various types of clusters: isolated spins, nn pairs, nn triads, ..., whose relative importance depends on $x$.\cite{Shapira:2002_JAP} The magnetic response of such small clusters can be easily calculated for given values of the exchange integral $J_{\text{nn}}$, temperature $T$, and magnetic field $H$. Possible interactions between more distant pairs of magnetic ions are neglected within this approach. Qualitatively, at low temperatures, $k_{\text{B}}T \ll |J_{\text{nn}}|$, at which the nn pairs are locked, the Curie constant is reduced ($C < C_0$) or enhanced ($C > C_0$) for antiferromagnetic interactions ($J_{\text{nn}}< 0$) and ferromagnetic interactions ($J_{\text{nn}}> 0$), respectively.

We are interested in the case of (Ga,Mn)N and (Ga,Mn)N:Si where, in general, both Mn$^{2+}$ ($S=5/2$) and Mn$^{3+}$ ($S=2$) ions are present with the concentrations $x_{\text{Mn}^{2+}}$ and $x_{\text{Mn}^{3+}}$, respectively ($x=x_{\text{Mn}^{2+}}+x_{\text{Mn}^{3+}}$). Here, we have ferromagnetically coupled Mn$^{3+}$--Mn$^{3+}$ and Mn$^{3+}$--Mn$^{2+}$ pairs as well as antiferromagnetically interacting Mn$^{2+}$--Mn$^{2+}$ pairs. To describe this case, we generalize the approach put previously forward for II-VI DMSs.\cite{Shapira:2002_JAP} The probability that a given spin, {\em e.g.}, a Mn$^{3+}$ ion, is in a cluster belonging to the configuration $r_{\text{cl}}$ is given by,
\begin{equation}
\label{eq:P_Mn2pMn3p}
P_{r_{\text{cl}}}=n_{r_{\text{cl}}}(x_{\text{Mn}^{2+}})^{n_{\text{Mn}^{2+}}}(x_{\text{Mn}^{3+}})^{n_{\text{Mn}^{3+}}-1}(1-x)^{\upsilon_{r_{\text{cl}}}},
\end{equation}
where $n_{\text{Mn}^{2+}}$  and $n_{\text{Mn}^{3+}}$ are the numbers of manganese ions in 2+ and 3+ charge states belonging to this cluster and $n_{r_{\text{cl}}}$, $\upsilon_{r_{\text{cl}}}$ are parameters taken from Refs.~\onlinecite{Shapira:2002_JAP} and \onlinecite{Shapira:2002_JAP_EPAPS}. Then one should take into account all possible combinations of arrangements of Mn$^{2+}$ ions within $r_{\text{cl}}$ and calculate the total spin $S_{\text{cl}}$ corresponding to the ground state (assuming that $|J_{\text{nn}}|\gg k_{\text{B}}T$). In the case of vanishing interactions between Mn$^{2+}$ ions we have to compute also the effective cluster size $n_{\text{cl}}$ (some of the ions may be disconnected from the initial cluster because of $J_{\text{Mn}^{2+}-\text{Mn}^{2+}}=0$ ). In this way we obtain the probability matrix $P_{S_{\text{cl}},n_{\text{cl}}}$ from which the Curie constant can be calculated,

\begin{equation}
\label{eq:CurieConstantCluster}
C=xN_0\frac{{(g\mu_{\text{B}})}^2}{3k_{\text{B}}}\sum_{S_{\text{cl}},n_{\text{cl}}} \frac{P_{S_{\text{cl}},n_{\text{cl}}}S_{\text{cl}}(S_{\text{cl}}+1)}{n_{\text{cl}}},
\end{equation}
where $N_0$ is the cation concentration.

In order to obtain the magnetization $M(T,H)$ of (Ga,Mn)N in the presence of interacting nn magnetic centers, we extend the previous model of a single substitutional Mn$^{3+}$ impurity in GaN (Refs.~\onlinecite{Stefanowicz:2010_PRBa,Wolos:2004_PRBb,Gosk:2005_PRB}) by considering a pair of Mn$^{3+}$ ions coupled by an exchange interaction $H(12)=-J_{\text{nn}}\vec{S}_1\vec{S}_2$. Then, the energy structure of such a pair can be described by the Hamiltonian
\begin{equation}
\label{eq:H_Pair}
H=H(1)+H(2)+H(12),
\end{equation}
where $H(i)$ ($i=1,2$) accounts for the single Mn$^{3+}$ ion ($L=2$, $S=2$) in GaN with the trigonal crystal field of the wurtzite structure and the Jahn-Teller
distortion taken into account (for details, see Ref.~\onlinecite{Stefanowicz:2010_PRBa}). In the coupling scheme employed, the base states for the pair are characterized by the set of quantum numbers $\mid m_{L_1},m_{S_1},m_{L_2},m_{S_2}\rangle$. The energy level scheme is calculated by a numerical diagonalization of the full $625 \times 625$ Hamiltonian (\ref{eq:H_Pair}) matrix, which allows to obtain an average magnetic moment of the Mn ion belonging to the pair (${\bm{m}}_{\text{pairs}}$). Assuming a random distribution of Mn ions over the cation sites (the hcp lattice) and allowing for the coupling between the nn spins we can then evaluate $M(T,H)$ at given $J_{\text{nn}}$ and $x$,
\begin{equation}
\label{eq:M_SingesPair}
\bm{M}=\mu_{\text{B}} xN_0 [\langle\bm{m}_{\text{singles}}\rangle P_{n_{\text{cl}}=1}+\langle\bm{m}_{\text{pairs}}\rangle(1-P_{n_{\text{cl}}=1})].
\end{equation}

\end{document}